\documentclass[10pt]{article}
\usepackage{graphicx}
\usepackage{xspace}
\usepackage{url}
\usepackage{booktabs}
\usepackage{wrapfig,rotating}
\usepackage{sidecap}
\usepackage{amsmath}
\usepackage{amssymb}
\usepackage{mathrsfs}
\usepackage{lineno}
\usepackage{color}

\def\Title#1{\begin{center} {\Large #1 } \end{center}}
\def\Author#1{\begin{center}{ \sc #1} \end{center}}
\def\Address#1{\begin{center}{ \it #1} \end{center}}

\newcommand\pubblock{\rightline{\begin{tabular}{l} Proceedings of the Second Annual LHCP\\ \pubnumber\\
         \pubdate  \end{tabular}}}

\newenvironment{Abstract}{\begin{quotation} \begin{center} 
             \large ABSTRACT \end{center}\bigskip 
      \begin{center}\begin{large}}{\end{large}\end{center} \end{quotation}}

\newenvironment{Presented}{\begin{quotation} \begin{center} 
             PRESENTED AT\end{center}\bigskip 
      \begin{center}\begin{large}}{\end{large}\end{center} \end{quotation}}

\def\Acknowledgements{\bigskip  \bigskip \begin{center} \begin{large}
             \bf ACKNOWLEDGEMENTS \end{large}\end{center}}

\def\beq{\begin{equation}}
\def\eeq#1{\label{#1}\end{equation}}
\def\eeqn{\end{equation}}

\def\beqa{\begin{eqnarray}}
\def\eeqa#1{\label{#1}\end{eqnarray}}
\def\eeqan{\end{eqnarray}}

\let\bar=\overbar

\def\Dslash{\not{\hbox{\kern-4pt $D$}}}
\def\dslash{\not{\hbox{\kern-2pt $\del$}}}

\def\msb{{\bar{\ssstyle M \kern -1pt S}}}

\usepackage{color}
\usepackage{xspace}

\newcommand\pubnumber{FERMILAB-CONF-14-339-E}

\newcommand\pubdate{\today}
\newcommand{\ljets}{$\ell +$jets\xspace}
\newcommand{\dzero}     {D0\xspace}

\newcommand{\ttbar}{\ensuremath{t\bar{t}}\xspace}
\newcommand{\mm}       {\mathrm}
\newcommand{\mTT}{\ensuremath{m_{t\bar{t}}}\xspace}

\newcommand{\afb}{\ensuremath{A_{\mbox{{\footnotesize FB}}}^{t\bar{t}}}\xspace}
\newcommand{\afbl}{\ensuremath{A_{\mbox{{\footnotesize FB}}}^{\mm{lep}}}\xspace}
\newcommand{\ac}{\ensuremath{A_{\mbox{{\footnotesize C}}}^{t\bar{t}}}\xspace}
\newcommand{\acl}{\ensuremath{A_{\mbox{{\footnotesize C}}}^{\mm{lep}}}\xspace}
\newcommand{\mcatnlo}   {\textsc{mc@nlo}\xspace}

\def\affiliation{
On behalf of the ATLAS, CDF, CMS and D0 Experiments, \\
Particle Physics Department \\
Fermilab, Batavia ,IL 60510, U.S.A }

\begin{document}

\large
\begin{titlepage}
\pubblock

\vfill
\Title{ Properties of the top quark }
\vfill

\Author{ Andreas W. Jung }
\Address{\affiliation}
\vfill
\begin{Abstract}
Recent measurements of top-quark properties at the LHC and the Tevatron are presented. Most recent measurements of the top quark mass have been carried out by CMS using $19.7/$fb of $\sqrt{s} = 8$ TeV data including the study of the dependence on event kinematics. ATLAS uses the full Run I data at $\sqrt{s} = 7$ TeV for a "3D" measurement that significantly reduces systematic uncertainties. \dzero employs the full Run II data using the matrix element method to measure the top quark mass with significantly reduced systematic uncertainties. Many different measurements of the top quark exist to date and the most precise ones per decay channel per experiment have been combined into the first world combination with a relative precision of 0.44\%. Latest updates of measurements of production asymmetries include the measurement of the \ttbar production asymmetry by \dzero employing the full Run II data set, by CMS and ATLAS (including the polarization of the top quark) employing both the full data set at $\sqrt{s} = 7$ TeV. CMS uses the full $\sqrt{s} = 8$ TeV data to measure the top quark polarization in single top production, the ratio ${\cal R}$ of the branching fractions ${\cal B}(t \rightarrow Wb) / {\cal B}(t \rightarrow Wq)$ and to search for flavor changing neutral currents. The results from all these measurements agree well with their respective Standard Model expectation.
\end{Abstract}
\vfill

\begin{Presented}
The Second Annual Conference\\
 on Large Hadron Collider Physics \\
Columbia University, New York, U.S.A \\ 
June 2-7, 2014
\end{Presented}
\vfill
\end{titlepage}
\def\thefootnote{\fnsymbol{footnote}}
\setcounter{footnote}{0}
%

\normalsize 


\section{Introduction}

The top quark is the heaviest known elementary particle and was discovered at the Tevatron $p\bar{p}$ collider in 1995 by the CDF and \dzero collaboration \cite{top_disc1,top_disc2} with a mass around $173~\mathrm{GeV}$. At the Tevatron the production is dominated by the $q\bar{q}$ annihilation process, while at the LHC the gluon-gluon fusion process dominates. The top quark has a very short lifetime, which prevents the hadronization process of the top quark. Instead bare quark properties can be observed. Measurements in the top quark sector are becoming highly precise nowadays, especially measurements of the top quark mass go well below 0.5\% in relative uncertainties.\\
The measurements presented here are performed using either the dilepton ($\ell \ell$) final state or the lepton+jets (\ljets) final state. Within the \ljets~final state one of the $W$ bosons (stemming from the decay of the top quarks) decays leptonically, the other $W$ boson decays hadronically. For the dilepton final state both $W$ bosons decay leptonically. The branching fraction for top quarks decaying into $Wb$ is almost 100\%. Jets originating from a $b$-quarks are identified ($b$-tagged) by means of multi-variate methods employing variables describing the properties of secondary vertices and of tracks with large impact parameters relative to the primary vertex.

\section{Top Quark Mass}
\label{toc:mass}

The presented measurements rely on different techniques in order to extract the top quark mass. The measurements either apply the leading order Matrix Element method based on an event-by-event probability, the Ideogram method based on an event likelihood, the template method comparing histograms of sensitive variables in data to simulations or alternative methods, such as endpoint- or $J/\psi$-method. All top quark mass measurements applying standard methods are dominated by systematic uncertainties and the most dominant ones are related to the $b$-jet energy scale (JES), the choice of the signal generator and the modeling of the hadronization and color reconnection effects. Furthermore there is an additional uncertainty, which originates from the implementation of the quark mass in the MC employed to measure the top quark mass.
\begin{figure}[ht]
    \centering \includegraphics[width=0.9\columnwidth]{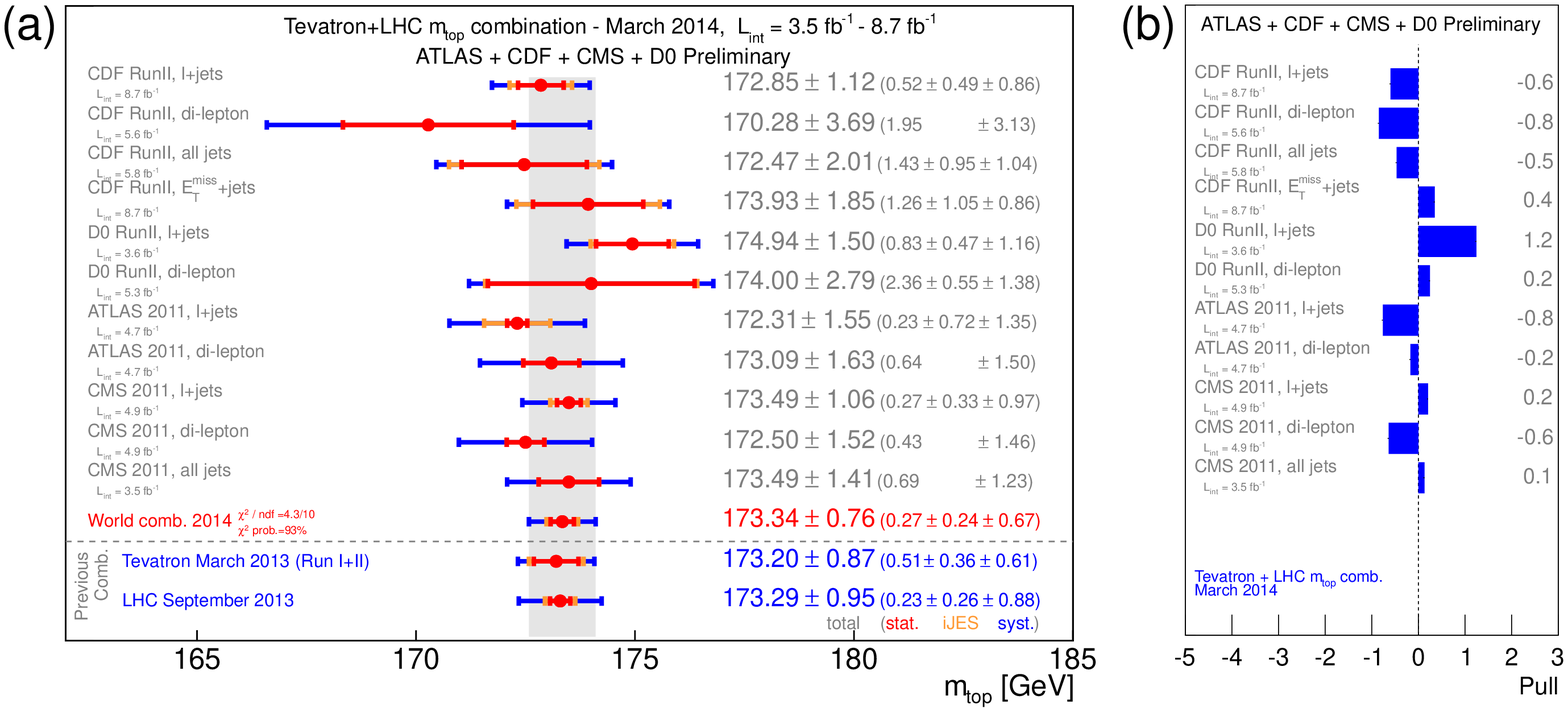}
  \protect\caption{\label{fig:world} Summary of (a) input measurements to the first LHC + Tevatron top quark mass combination compared to earlier combinations at LHC and Tevatron. Measurements are in good agreement with each other as indicated by the (b) pulls.}
 \end{figure}
This is a theoretical uncertainty on top of the experimental uncertainties and aims to answer whether the implemented mass definition is close to the pole mass or $\bar{MS}$ mass and which uncertainty is associated with the definition. Currently this uncertainty is of the order of $0.5$ to $1$ GeV. Strategies to overcome the limitations in terms of experimental and theoretical uncertainties are already pursued and will become more important for the upcoming run of the LHC. Measurements less sensitive to these uncertainties are, for example, employing multi-dimensional fits to extract the top quark mass or alternative methods with largely orthogonal correlations of experimental systematic uncertainties.\\
The first world combination of top quark mass measurements \cite{mt_world} by all four collaborations combines the most precise measurement per decay channel and per experiment. After careful and detailed study of the correlations of systematic uncertainties a total of 11 individual measurements are combined using the BLUE method. Figure \ref{fig:world}(a) shows the input measurements compared to the world average, Tevatron only and LHC only combinations. The input measurements are in good agreements as demonstrated by the pulls shown in Figure \ref{fig:world}(b). The combined top quark mass is $m_{t} = 173.34 \pm 0.76\,(\mm{stat.\,+\,sys.\,+\,JES})~\mm{GeV}$, corresponding to a total relative uncertainty of 0.44\%.\\
In the following the latest updates by the four collaborations on top quark mass measurements are discussed. CMS uses the full Run II data at $\sqrt{s} = 8$ TeV to measure the top quark mass in the \ljets decay channel \cite{cms_mt_kin}. After requiring exactly 2 $b$-tags a purity of 95\% is achieved allowing a precise measurement of the top quark mass and its dependence of various kinematic quantities. The observed dependencies appear since the method is not re-calibrated in each bin of a given quantity, which results in dependencies of the top quark mass seen in data and modeled properly by the MC. The quantities studied are sensitive to the models and MC tunes employed to study systematic uncertainties.
\begin{figure}[ht]
     \centering \includegraphics[width=0.85\columnwidth]{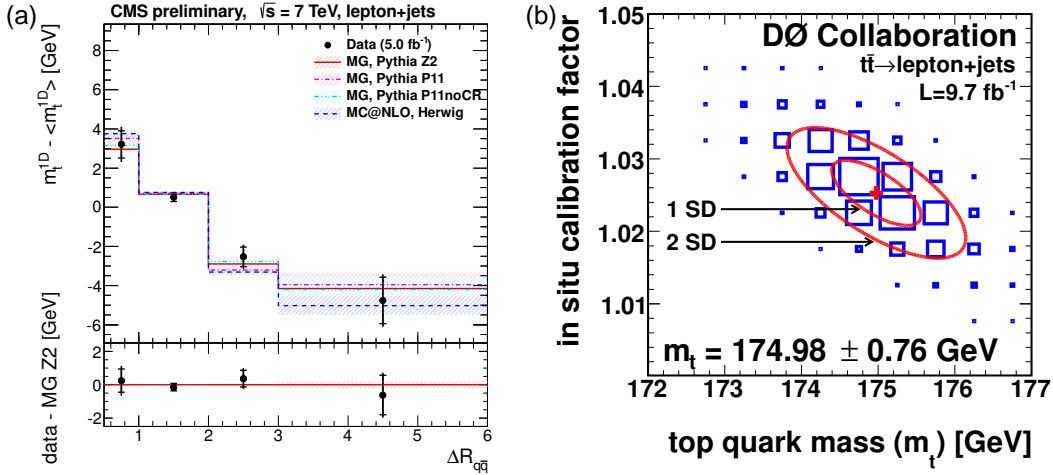}
  \protect\caption{\label{fig:mass_cms_d0} Kinematic dependence of the (a) top mass on $\Delta R_{q\bar{q}}$ by CMS compared to various models sensitive to color reconnection, more details in the text. The (b) two-dimensional likelihood as a function of the top quark mass and the in-situ calibration factor as measured by \dzero.}
 \end{figure}
Figure \ref{fig:mass_cms_d0}(a) shows that dependence as a function of $\Delta R_{q\bar{q}}$. The distribution is sensitive to the modeling of color reconnections but currently all models are in agreement with the various tunes indicating that more statistics is needed to understand the interplay. The measurement yields $m_{t} = 172.04 \pm 0.77\, (\mm{stat.\,+\,sys.\,+\,JES})~\mm{GeV}$ corresponding to a relative uncertainty of 0.45\%. ATLAS employs the full $\sqrt{s} = 7$ TeV data for a three-dimensional template method to determine $m_t$, jet energy scale factor (JSF) for light quarks and the JSF for $b$-quarks \cite{atlas_mt_3d}. Variables sensitive to the top mass and the JSFs are $m_t^{\mm{reco}}$, $m_W^{\mm{reco}}$ and $R_{\ell b}^{\mm{reco}}$ depends more strongly on the jet energy scale related to $b$-jets. The measurement yields $m_{t} = 172.31 \pm 0.75\, (\mm{stat.\,+\,JSF\,+\,bJSF}) \pm 1.35\, (\mm{sys.})~\mm{GeV}$. In the case of \dzero, the most precise measurement is done in the \ljets decay channel \cite{d0_mass} employs the so-called matrix element method (ME), which calculates an event-by-event probability to match the \ttbar final state in the \ljets decay channel to the observed reconstructed objects taking into account detector resolutions. The transfer function relates the probability density of measured quantities to the partonic quantities. As one of the $W$ bosons decays hadronically, a constraint on the $W$ mass can be used to fit the jet energy scale in-situ and derive an additional calibration factor. The measurement uses $9.7/$fb and is currently one of the most precise mass measurements by all four collaborations. Figure \ref{fig:mass_cms_d0}(b) shows the two-dimensional likelihood as a function of the top quark mass and the in-situ calibration factor. It yields a mass of $m_{t} = 174.98 \pm 0.41\, (\mm{stat.}) \pm 0.64\,(\mm{sys.\,+\,JES})~\mm{GeV}$, corresponding to a total relative uncertainty of 0.43\%. The CDF measurement in the \ljets decay channel yields $m_{t} = 172.85 \pm 1.12\,(\mm{tot.})~\mm{GeV}$ \cite{cdf_mass}, corresponding to a total relative uncertainty of 0.65\%.\\
The latest measurement of the mass difference between the top and anti-top quark is carried out by ATLAS and yields $\Delta m_t = 0.67 \pm 0.61\,(\mm{stat.}) \pm 0.41\,(\mm{sys.})$ GeV \cite{atlas_deltaM}. All measurements of the mass difference of the top and anti-top quark by CDF, CMS and \dzero \cite{cdf_deltaM, cms_deltaM, d0_deltaM}, as well as a search for Lorentz invariance violation by \dzero \cite{d0_Liv}, are consistent with CPT invariance.\\
Direct measurements of the top quark mass are becoming ever more precise and provide a stringent self-consistency test of the SM by correlating $m_t$ versus $m_W$. Together with the measurement of the mass of the recently discovered Higgs boson \cite{higgs1, higgs2} this is a strong self-consistency test of the SM \cite{gfitter}. Furthermore the stability of the electroweak vacuum can be studied and currently the preferred experimental range indicates that the vacuum is meta-stable \cite{vacuum}. The current measurements and the theoretical extrapolation seem to indicate that the vacuum is meta-stable, and more measurements are needed to fully understand this relation.

\section{Top quark production asymmetries}
\label{toc:angular}
\begin{figure}[ht]
  \centering
  \includegraphics[width=0.95\columnwidth,angle=0]{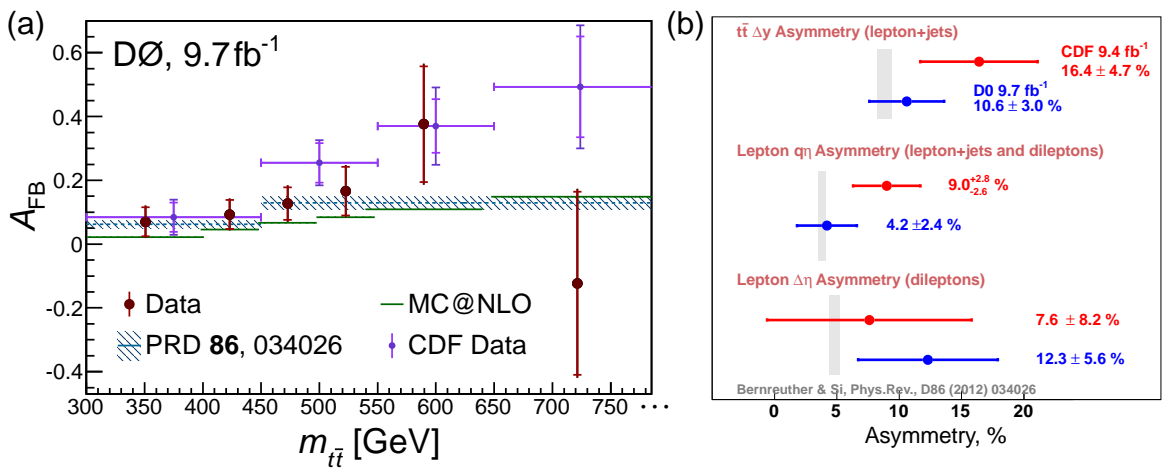}
\protect\caption{\label{fig:cdf_mttafb} The (a) \afb at parton level as a function of the invariant mass of the \ttbar pair $M_{t\bar{t}}$ as measured by CDF and \dzero compared to the predicted dependency by NLO QCD $\oplus$ electroweak corrections \cite{bernSi} or \mcatnlo. Summary of (b) \afb and \afbl measurements at the Tevatron. For \afbl these are results of the combination of results in the \ljets and dilepton decay channel.}
\end{figure}

The different initial state makes measurements of angular correlations in $t\bar{t}$ events, such as production asymmetries, complementary between the Tevatron and the LHC. Experimentally, there are two approaches to measure these asymmetries: Either top quarks are fully reconstructed using a kinematic reconstruction or only a final-state particle, e.g. a lepton (`lepton-based asymmetries') is reconstructed. The latter avoids the reconstruction of top-quarks, which is usually more affected by detector resolution and migration effects. The forward-backward asymmetry \afb at the Tevatron measures $\Delta y = y_t - y_{\bar{t}}$, whereas the charge asymmetry \ac at the LHC measures $\Delta |y| = |y_t| - |y_{\bar{t}}|$ and employing these the production asymmetries are defined as
\begin{equation}
A_{\mbox{{\footnotesize FB}}}^{t\bar{t}} = \dfrac{N(\Delta y >0) - N(\Delta y <0)}{N(\Delta y >0) + N(\Delta y <0)},\,\,\mm{and}\,\,A_{\mbox{{\footnotesize C}}}^{t\bar{t}} = \dfrac{N(\Delta |y| >0) - N(\Delta |y| <0)}{N(\Delta |y| >0) + N(\Delta |y| <0)},\,\,\mm{respectively.}
\end{equation}
As mentioned above an additional observable is given by the lepton-based asymmetries, which are similarly defined only that instead of top quark rapidities, the rapidities of the decay leptons are used to measure the production asymmetries.\\
\begin{figure}[th]
  \centering
  \includegraphics[width=0.80\columnwidth,angle=0]{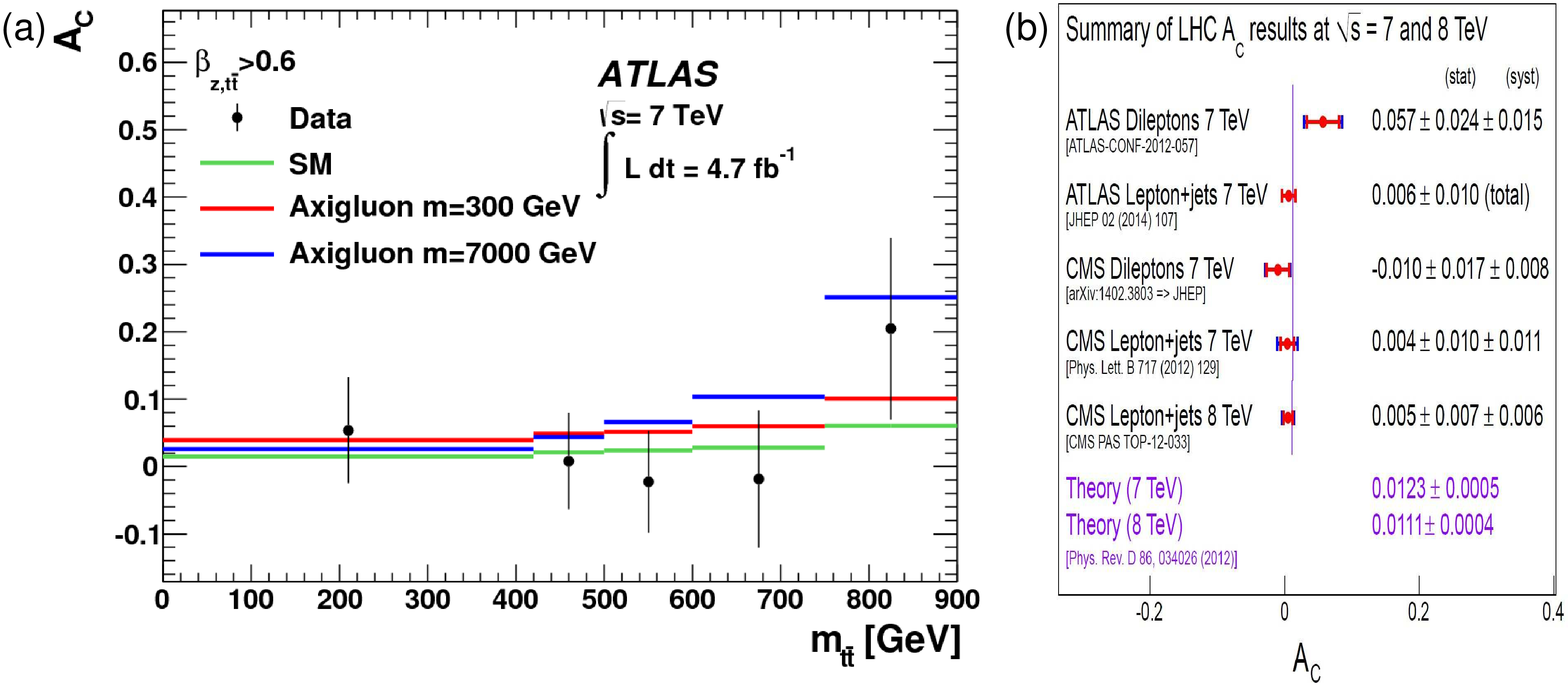}
\protect\caption{\label{fig:lhc_ac} The (a) \ac at parton level as a function of (a) $M_{t\bar{t}}$ for $\beta_z^{t\bar{t}} > 0.6$ and a (b) summary of \ac measurements at the LHC compared to the predictions at $\sqrt{s} = 7$ or $8$ TeV.}
\end{figure}
One of the latest measurements of \ac at ATLAS uses the full $\sqrt{s} = 7$ TeV data in the \ljets decay channel with at least 1 $b$-tag \cite{atlas_ac}. In addition to the inclusive \ac also the kinematic dependencies of \ac from \mTT and $\beta_z^{t\bar{t}}$ are measured and found to be in agreement with the SM predictions as shown in Figure \ref{fig:lhc_ac}(a). Given the large uncertainties the data are also in agreement with two exemplary beyond the SM (BSM) models implementing contributions of axi gluons. The inclusive measurement yields \ac$=0.006 \pm 0.010$, and is in agreement with the theory prediction of \ac$=0.0123 \pm 0.0005$.\\
The latest update by CMS measures \ac also employing the full $\sqrt{s} = 7$ TeV data, but the dilepton decay channel with at least 1 $b$-tag \cite{cms_ac}. Top quarks are reconstructed using the analytical matrix weighting technique. The measurement also includes the kinematic dependency of \ac from \mTT and in addition the measurement of \ac using decay leptons, with results being in agreement to the SM predictions. The inclusive measurements yield \ac$=-0.010 \pm 0.019$ and \acl$=0.009 \pm 0.012$ compared to the theoretical prediction of \ac$=0.0123 \pm 0.0005$ and \acl$=0.0070 \pm 0.0003$, respectively.\\
CDF uses data corresponding to $9.4~\mathrm{fb^{-1}}$ of integrated luminosity and employs a kinematic reconstruction to reconstruct the \ttbar final state in the \ljets decay channel \cite{cdf_ttbar_afb}. CDF measures an inclusive asymmetry of \afb$=0.164 \pm 0.045$ (stat. + syst.) at the parton level compared to the SM prediction of \afb$=0.088 \pm 0.005$ (NLO QCD $\oplus$ electroweak corrections) \cite{bernSi}. In addition the kinematic dependency of \afb is extracted, by measuring $\Delta y$ in bins of $M_{t\bar{t}}$, as shown in Fig.~\ref{fig:cdf_mttafb}(a).
The CDF results show a dependence on $M_{t\bar{t}}$, which is different from the SM expectation by 2.4 standard deviations.\\
\dzero uses the full Run II data, corresponding to $9.7~\mm{fb^{-1}}$ of integrated luminosity \cite{d0_ttbar_afb}, and also fully reconstructs the \ttbar final state using a kinematic reconstruction. The measurement in the \ljets decay channel
results in an inclusive asymmetry of \afb$=0.106 \pm 0.030$ (stat. + syst.) at the parton level. The result is compatible with the SM and results by CDF. \dzero does not see an indication for a strong \mTT dependency beyond the one expected by the SM as shown in Fig.~\ref{fig:cdf_mttafb}(a). It should be noted that very recently predictions at NNLO pQCD by Mitov et al.~became available (presented at the CKM14 conference) with a predicted value at NNLO including electroweak corrections of \afb$\approx$ 10\%, which are in agreement with the Tevatron data. However, differential predictions of \afb in \mTT are needed to fully understand the picture.\\
The \dzero result in terms of the lepton-based asymmetries in the \ljets channel is \afbl$= 0.047 \pm 0.026$ (stat. + syst.) at the parton level \cite{d0_leptonic_afb} and in the dilepton channel the corresponding measurement is \afbl$ = 0.044 \pm 0.039$ (stat. + syst.), whereas the dilepton asymmetry is measured to be $A^{\ell \ell} = 0.123 \pm 0.056$. A summary of \afb and \afbl measurements at the Tevatron is given in Figure \ref{fig:cdf_mttafb}(b). It is interesting to note that the ratio of the two lepton-based asymmetries in the dilepton channel shows a deviation from the SM prediction of about two standard deviations.\\
CDF employed data corresponding to up to $9.4~\mm{fb^{-1}}$ of integrated luminosity and performed a combination of \afbl  measurements. After combining results from \ljets and dilepton channels \afbl is $0.09 ^{+0.028}_{-0.026}$ \cite{CDF-CONF-2013-11035}, see Figure \ref{fig:cdf_mttafb}(b).\\
Currently, the results from the LHC are not yet significant enough in order to make a precise statement on the agreement with the SM. However, some BSM models are strongly disfavored by the LHC data, while others are still compatible. For measurements of \afb the deviations from the SM predictions got smaller with the new \dzero measurement employing the full data set, but are still higher than the SM predictions. CDF results with the full data set are showing deviations at the two s.d.~level. It should be noted that the individual results on \afb and \afbl employ the full data recorded by CDF and \dzero and studies on combinations are currently ongoing.

\section{Top quark polarization}
\label{toc:polarization}

\begin{wrapfigure}{r}{0.545\textwidth}
\centerline{\includegraphics[width=0.505\textwidth]{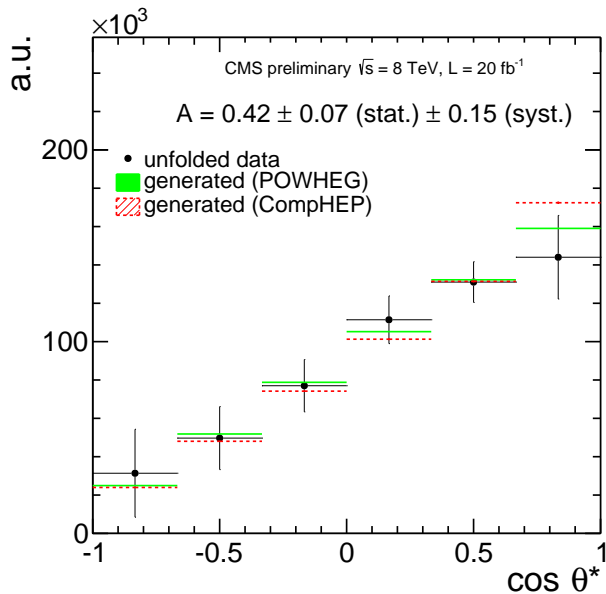}}
\caption{\label{fig:cmsljets7tev} The unfolded $\cos \theta^*$ distribution in measured in $t$-channel single top quark events used to measure the polarization of the top quark.}
\end{wrapfigure}%
Measurements of the polarization of the top quark in \ttbar production can provide hints on contributions of new physics since no polarization is expected in the SM and new physics can polarize top quarks. The latest measurement by ATLAS of the top quark polarization assumes that the polarization is either introduced by $CP$ conserving ($CPC$) or violating processes ($CPV$) \cite{atlas_pol}. With the spin analyzing power $\alpha_{l}$ the measurement assuming $CPC$ processes yields \mbox{$\alpha_{l}P_{CPC} = -0.035 \pm 0.014\,(\mm{stat.}) \pm 0.037\,(\mm{sys.})$} \\ and the measurement assuming $CPV$ yields $\alpha_{l}P_{CPV} = 0.020 \pm 0.016\,(\mm{stat.}) ^{+0.013}_{-0.017}\,(\mm{sys.})$, both are in agreement with the SM expectation of negligible polarization. Good agreement with the SM is also observed by earlier measurements in CMS \cite{cms_spin_pol} and \dzero \cite{d0_pol}.\\
In contrast to \ttbar production, where negligible top quark polarization is expected, in the production of single top quarks the top quarks are expected to be polarized in the SM. CMS employed the full data set at $\sqrt{s} = 8$ TeV to select single top quark events in the $t$-channel \cite{cms_pol_stop}. The polarization agrees with SM expectations and is measured to be \mbox{$P_t = 0.82 \pm 0.12\,(\mm{stat.}) \pm 0.32\,(\mm{sys.})$}.

\section{Branching fraction ${\cal R}$ and rare decays of the top quark}
\label{toc:r_rare}
Measurements of the branching fraction ${\cal R} = {\cal B}(t \rightarrow Wb) / {\cal B}(t \rightarrow Wq)$ and deviations in ${\cal R}$ from the SM expectation could indicate contributions of new physics, such as a charged Higgs. The latest update is carried out by CMS employing the full data set at $\sqrt{s} = 8$ TeV \cite{cms_r}. Events are selected in the dilepton decay channel separated by the number of jets and $b$-tags for the three dilepton channels of $ee$, $\mu\mu$ and $e\mu$. The measurement is currently the most precise measurement of ${\cal R}$ and yields $1.014 \pm 0.003\,(\mm{stat.}) \pm 0.032\,(\mm{sys.})$ with a lower limit of ${\cal R} > 0.955$ at 95\% confidence level (CL). In addition a lower limit for $V_{tb}$ is extracted as well and it is $0.975$ at 95\% CL. Other measurements are also in agreement with the SM and no hints for contribution of new physics are seen \cite{d0_r,cdf_r}.\\
Another probe to identify contributions of new physics are searches for processes involving flavor changing neutral currents (FCNC). Such processes are highly suppressed in the SM but large enhancements are possibly in many models of new physics. One of the latest updates in this area is the search for FCNC in \ljets final state with additional 2 leptons originating from the decay of the $Z$ boson done by CMS \cite{cms_fcnc}. This search sets various limits on a variety of FCNC processes, such as ${\cal B}(t \rightarrow ug)$ and ${\cal B}(t \rightarrow cg)$, but no indications of FCNC are observed. Similar searches have been performed earlier also at ATLAS \cite{atlas_fcnc} and \dzero \cite{d0_fcnc} and show no indication of FCNC.

\section{Conclusions}

Various recent measurements of top quark properties at the LHC and at the Tevatron are discussed. Direct measurements of the top quark mass are becoming ever more precise and provide a stringent self-consistency test of the SM and new insights into the question of the stability of the electroweak vacuum. Measurements of production asymmetries at the LHC are not yet significant enough in order to make a precise statement on the agreement with the SM. However, some BSM models are strongly disfavored by the LHC data, while others are still compatible with data. For measurements of \afb and \afbl at the Tevatron the deviations from the SM predictions got smaller with the new \dzero measurement employing the full data set. Studies on combinations of \afb and \afbl at the Tevatron are currently ongoing. Top quark polarization has been observed in single top quark production and is in agreement with the SM. All of the presented results in terms are in good agreement with the Standard Model expectations and do not show any hints for new physics.
\Acknowledgements
The author thanks the organizers of the LHCP 2014 conference for the invitation and for the hospitality of the conference venue.

\end{document}